\documentclass[10.5pt,a4paper,oneside]{extarticle}
\usepackage[T1]{fontenc}
\usepackage[left=25mm,right=20.9mm,top=3.75cm,bottom=3.75cm,headsep=0.25cm,footskip=0.25cm]{geometry}

\usepackage[square, comma, sort&compress, numbers]{natbib}
\usepackage[english]{babel}
\usepackage[affil-it]{authblk}
\usepackage{etoolbox}
\usepackage{lmodern}
\usepackage{setspace}
\usepackage{amssymb,amsthm,amsmath}
\usepackage{tikz}
\usepackage{listings}
\usepackage{color}
\usepackage{caption}
\usepackage{placeins}
\usepackage{booktabs}
\usepackage{enumerate}
\usepackage{graphicx}
\usepackage{titlesec}
\usepackage{mathptmx}
\usepackage{subfigure}
\usepackage{anyfontsize}
\usepackage{t1enc}
\usepackage[utf8]{inputenc}
\usepackage{fancyhdr}
\usepackage{float,xspace}
\usepackage{scalefnt}
\usepackage{soul}
\usepackage{tabulary}
\usepackage[colorlinks=true,linkcolor=blue]{hyperref}
\usepackage{textcase}

\usepackage[colorinlistoftodos]{todonotes}

\titlespacing*{\section}
{0pt}{12pt}{0pt}
\titlespacing*{\subsection}
{0pt}{12pt}{0pt}
\titlespacing*{\subsubsection}
{0pt}{12pt}{0pt}
\titleformat{\section}
  {\huge\sffamily\Large\bfseries\color{black}}
  {\thesection}{1em}{}
    \titleformat{\subsection}[block]{\bfseries}{\thesubsection}{.5em}{}
    \titleformat{\subsubsection}[block]{\bfseries}{\thesubsubsection}{.5em}{}
\titleformat{\section}{\fontsize{12}{19}\bfseries}{\thesection}{1em}{}
\usepackage{mathptmx}
\makeatletter

\patchcmd{\@maketitle}{\LARGE \@title}{\fontsize{14}{19.2}\selectfont\@title}{}{}

\title
{
	\vspace{-5cm}
	\begin{minipage}{\textwidth}	
	\hspace{-20pt}\includegraphics[width=9.5cm]{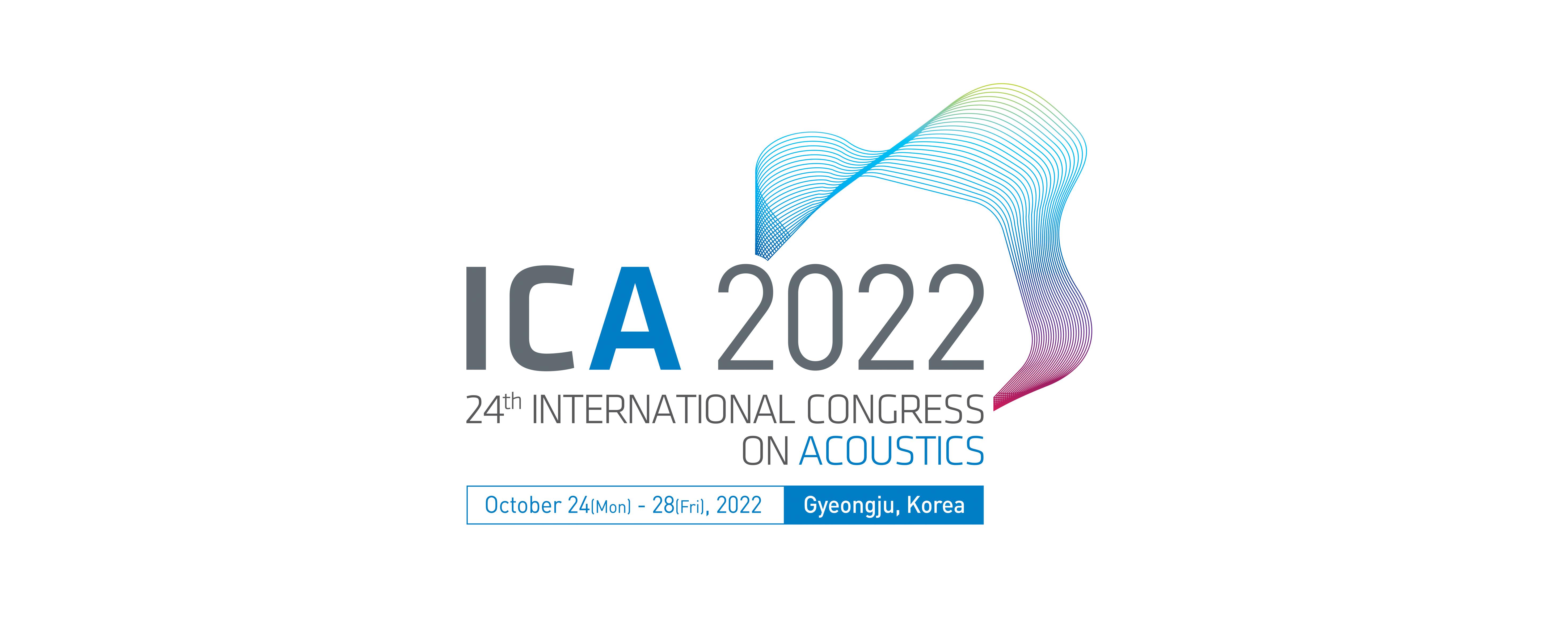}
	\hspace{-65pt}\includegraphics[width=10.2cm]{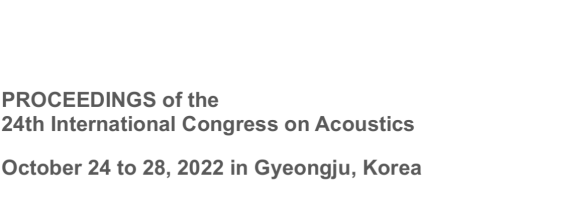}
	\end{minipage}
	\includegraphics[width=16.5cm]{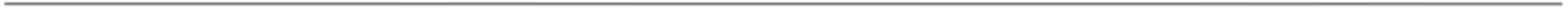}\\[0.5cm] \textbf{Implementation of Multi-channel Active Noise Control based on Back-propagation Mechanism}
	\author{Zhengding LUO, Dongyuan SHI, Junwei JI, Woon-Seng GAN}
  	\affil{School of Electrical \& Electronic Engineering, Nanyang Technological University, Singapore}
}
\date{}

\begin{document}
\setcitestyle{round} 

\clearpage
\setcounter{page}{1}
\maketitle
\thispagestyle{empty}
\fancypagestyle{empty}
{	
	\fancyhf{} \fancyfoot[R]
	{
		\vspace{-2cm}\includegraphics[width=17cm]{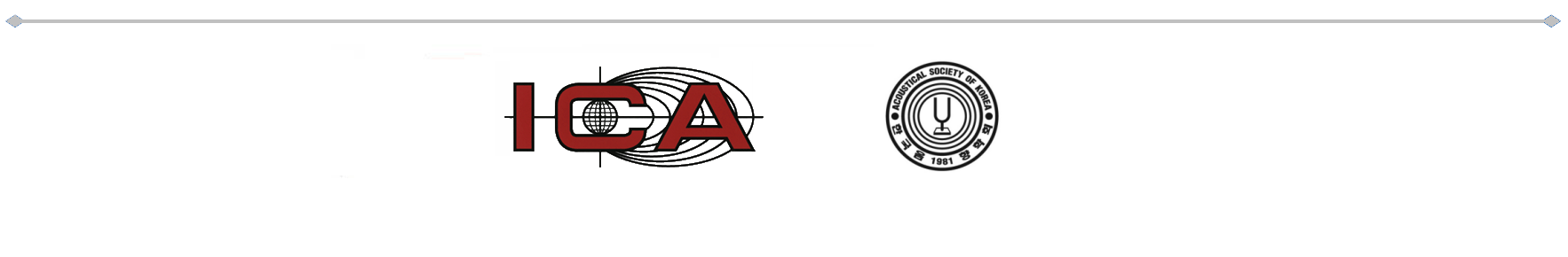}
	}
}

\subsection*{\fontsize{10.5}{19.2}\uppercase{\textbf{Abstract}}}
{\fontsize{10.5}{60}\selectfont Active noise control (ANC) systems can efficiently attenuate low-frequency noises by introducing anti-noises to combine with the unwanted noises. In ANC systems, the filtered-x least mean square (FxLMS) and filtered-X normalized least-mean-square (FxNLMS) algorithm are well-known algorithms for adaptively adjusting control filters. Multi-channel ANC systems are typically required to attenuate unwanted noises in a large space. However, open-source implementations of the multi-channel FxLMS (McFxLMS) and multi-channel FxNLMS (McFxNLMS) algorithm continue to be scarce. Therefore, this paper proposes a simple and effective implementation approach of the McFxLMS and McFxNLMS algorithm. Motivated by the back-propagation process during neural network training, the McFxLMS and McFxNLMS algorithm can be implemented via automatic derivation mechanism. We implemented the two algorithms using the automatic derivation mechanism in PyTorch and made the source code available on GitHub. This implementation method can improve the practicality of multi-channel ANC systems, which is expected to be widely used in ANC applications.}

\noindent{\\ \fontsize{11}{60}\selectfont Keywords: Active noise control, Multi-channel ANC, Implementation of McFxLMS and McFxNLMS} 

\fontdimen2\font=4pt

\section{\uppercase{Introduction}}
Passive noise control techniques such as enclosures, barriers, and silencers are difficult to suppress undesired noises at low frequencies \cite{1}. An active noise control (ANC) system is achieved by introducing a canceling “anti-noise” wave through an appropriate array of secondary sources \cite{2}. The anti-noise is of equal amplitude and opposite phase compared to the disturbance \cite{3}. In practice, ANC systems can efficiently attenuate low-frequency noises, whereas passive methods are either ineffective or tend to be very expensive or bulky \cite{13}. Therefore, ANC techniques have been widely employed in some commercial products including headphones, mobile phones and automobiles etc.

However, there is a performance degradation of single-channel ANC systems in spatial noise environments \cite{4}. In such scenarios, multi-channel ANC systems with multiple controllers, loudspeakers and microphones can be introduced to improve the noise reduction performance \cite{5}. Some of the best-known applications are the control of exhaust boom noises in automobiles \cite{6}, earth-moving machines \cite{7}, and the control of propeller-induced noises in flight cabin interiors \cite{8}. Other noise control applications like vibration control in complex mechanical structures also require multiple channels \cite{14}.

A general multiple-channel ANC system is shown in Figure \ref{Fig 1}, which consists of $J$ reference microphones, $K$ canceling loudspeakers, and $M$ error microphones. In the system, the active noise controller is updated based on the reference signals and error signals, sensed by the reference microphones and error microphones respectively, so that the generated anti-noise can suppress the disturbance at the error microphones. The coefficients of ANC controllers can be updated through adaptive algorithms to minimize the sum of the error signals \cite{9}.

Adaptive algorithms such as the filtered-x least mean square (FxLMS) \cite{10} and filtered-X normalized least-mean-square (FxNLMS) \cite{17} are commonly used ANC algorithms. Nevertheless, the adaptive optimization of control filter coefficients in multi-channel ANC systems is difficult to implement in practice \cite{15}. It must take into account the interactions between all microphones and loudspeakers to provide the optimal cancelling signals \cite{16}. Compared with single-channel ANC systems, making iterative changes to the control filter coefficients becomes much more complex in multi-channel ANC systems.

Due to the difficulty of implementation, open-source codes of the multi-channel FxLMS (McFxLMS) and multi-channel FxNLMS (McFxNLMS) algorithm are still lacking \cite{12}. To tackle the limitation, the McFxLMS and McFxNLMS algorithm are implemented via the automatic derivation mechanism in this paper. Since the automatic derivation can update weights given a loss function, the back propagation of McFxLMS and McFxNLMS can be automatically achieved without resorting to extra-human efforts. The implementation code is based on the PyTorch tool \cite{11} and available at \href{https://github.com/ShiDongyuan/Multichannel_FxLMS_python_code}{GitHub website}. Owing to the simple implementation method, changing the number of microphones and loudspeakers in multiple-channel ANC systems is more convenient, so that the practicality is improved.

\section{\uppercase{Forward Propagation and Back Propagation}}

\begin{figure}[tp]
\setlength{\abovecaptionskip}{0.cm}
\setlength{\belowcaptionskip}{-0.cm}
\centering
\centerline{\includegraphics[width=0.6\linewidth]{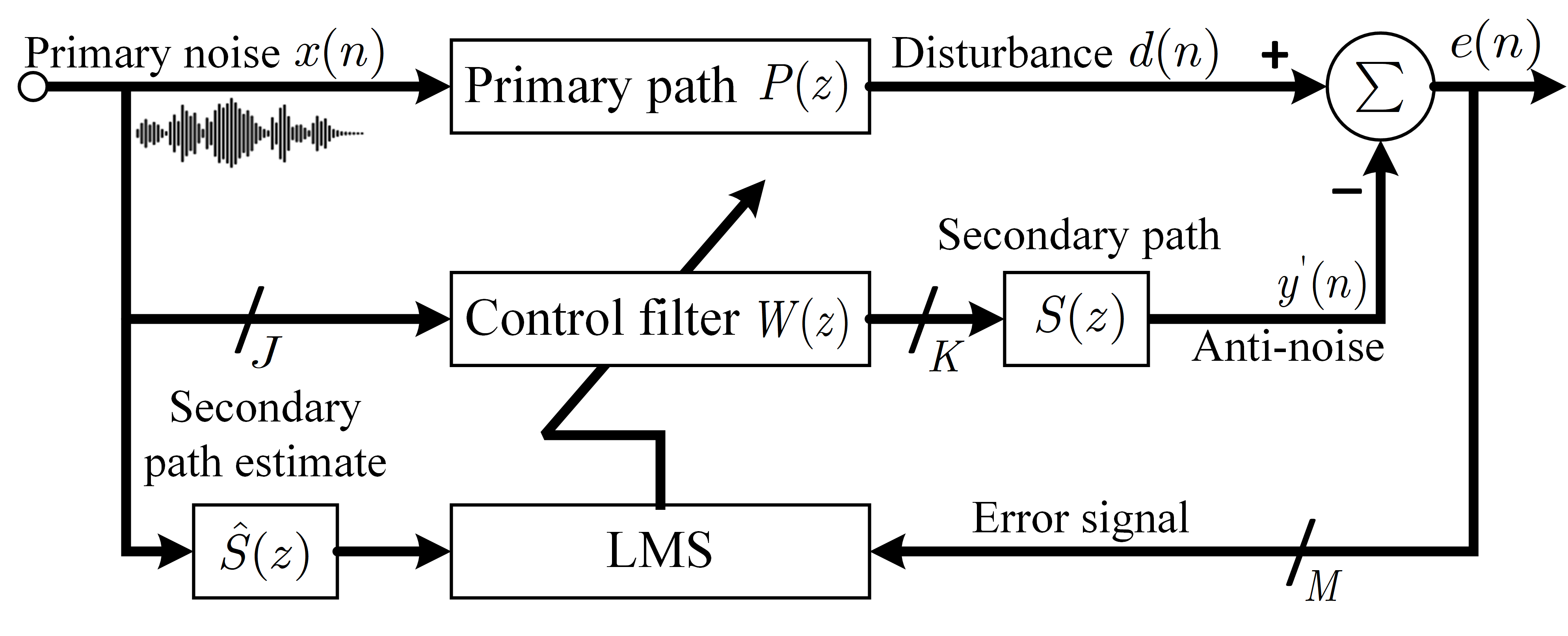}}
\caption{Block diagram of a $J \times K \times M$ multi-channel ANC system using the McFxLMS algorithm.}
\label{Fig 1}
\end{figure}

A feedforward multiple-channel ANC system using the McFxLMS algorithm is illustrated in Figure \ref{Fig 1}. The McFxLMS algorithm is used to update the active noise controller. The $J \times K \times M$ multiple-channel ANC system includes $J$ reference microphones, $K$ canceling loudspeakers, and $M$ error microphones. For the purpose of easily implementing adaptive algorithms in PyTorch, we split the adaptive process into two phases: forward propagation and back propagation. Then, the McFxLMS algorithm is described in detail as an example.

\subsection{Forward propagation}
\begin{figure}
\centering
\includegraphics[width=11cm]{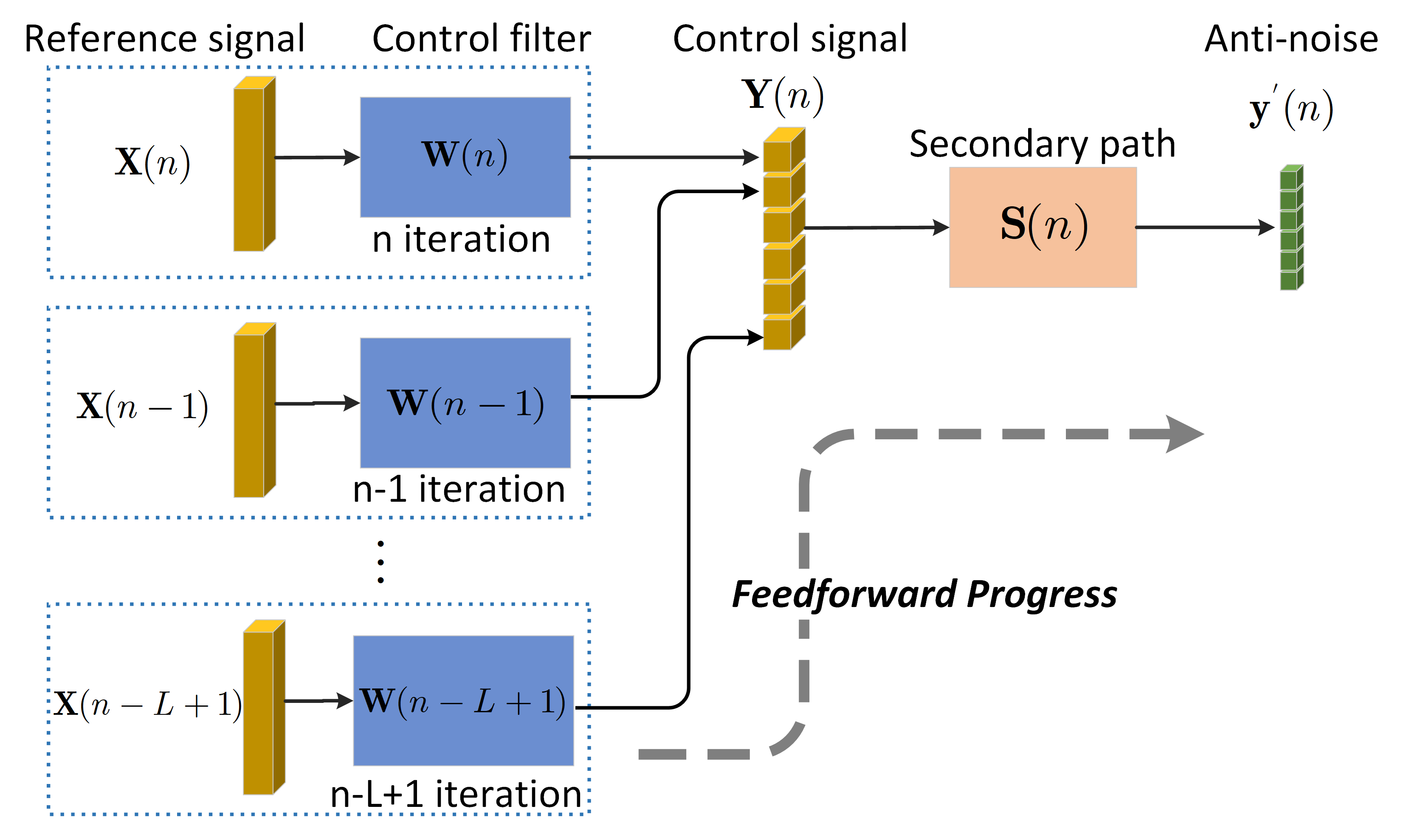}
\caption{The forward propagation of the McFxLMS algorithm.}
\label{Fig 2}
\end{figure}

In forward propagation, as shown in Figure \ref{Fig 2}, the input component of the McFxLMS algorithm at its $n$-th iteration is the reference vector $\mathbf{X}(n)$ given by
\begin{equation}
    \mathbf{X}(n)=\begin{bmatrix}
    \mathbf{x}^\mathrm{T}_1(n)&\cdots & \mathbf{x^\mathrm{T}}_j(n) &\cdots & \mathbf{x^\mathrm{T}}_J(n)
    \end{bmatrix}^\mathrm{T}\in\mathbb{R}^{JN\times 1},
\end{equation}
where $\mathrm{T}$ denotes the transpose operation. The control filter length is $\mathrm{N}$ taps. The reference signal vector picked up by the $j$-th microphone can be expressed as
\begin{equation}
    \mathbf{x}_j(n) =\begin{bmatrix}
    x_j(n) & x_j(n-1) &\cdots & x_j(n-N+1)
    \end{bmatrix}^\mathrm{T}\in\mathbb{R}^{N\times 1}.
\end{equation}
The control filter matrix is represented as
\begin{equation}
    \mathbf{W}(n)=\begin{bmatrix}
    \mathbf{w}^\mathrm{T}_{11}(n) &\cdots & \mathbf{w}^\mathrm{T}_{1J}(n)\\
    \mathbf{w}^\mathrm{T}_{21}(n) &\cdots & \mathbf{w}^\mathrm{T}_{2J}(n)\\
    \vdots                        &\ddots & \vdots                        \\
    \mathbf{w}^\mathrm{T}_{K1}    &\cdots & \mathbf{w}^\mathrm{T}_{KJ}(n)
    \end{bmatrix}\in\mathbb{R}^{K\times JN },
\end{equation}
The element $\mathbf{w}_{kj}(n)$ in control filter matrix means the control filter from the $j$th reference to the $k$th output and can be denoted as
\begin{equation}
    \mathbf{w}_{kj}(n)=\begin{bmatrix}
    w^{(1)}_{kj}(n) & w^{(2)}_{kj}(n) & \cdots & w^{(N)}_{kj}(n)
    \end{bmatrix}^\mathrm{T} \in\mathbb{R}^{N\times 1}.
\end{equation}

Hence, the control signal $\mathbf{y}(n)$ can be obtained from
\begin{equation}
    \mathbf{y}(n)=\mathbf{W}(n)\mathbf{X}(n) \in\mathbb{R}^{K\times 1},
\end{equation}
and it can also be expressed as
\begin{equation}
    \mathbf{y}(n) = \begin{bmatrix}
    y_1(n) & \cdots & y_k(n) & \cdots & y_K(n)
    \end{bmatrix} \in\mathbb{R}^{K\times 1}.
\end{equation}
As Figure \ref{Fig 2} illustrates, the overall control signal vector is formed by stacking the above control signal from $n-L+1$ iteration to $n$ iteration as
\begin{equation}
    \mathbf{y}_k(n) =\begin{bmatrix}
    y_k(n) & y_k(n-1) &\cdots & y_k(n-L+1)
    \end{bmatrix}^\mathrm{T}\in\mathbb{R}^{L\times 1},
\end{equation}
\begin{equation}
    \mathbf{Y}(n)=\begin{bmatrix}
    \mathbf{y}^\mathrm{T}_1(n)&\cdots & \mathbf{y^\mathrm{T}}_k(n) &\cdots & \mathbf{y^\mathrm{T}}_K(n)
    \end{bmatrix}^\mathrm{T}\in\mathbb{R}^{KL\times 1},
\end{equation}
in which $y_k(n)$ stands for the $k$-th control signal to drive the corresponding secondary source. $L$ means the length of secondary path.

Subsequently, the anti-noise vector is obtained from 
\begin{equation}
    \mathbf{y}^\prime(n)=\mathbf{S}(n)\mathbf{Y}(n) \in\mathbb{R}^{M\times 1},
\end{equation}
where the secondary path matrix is expressed as 
\begin{equation}
    \mathbf{S}(n)=\begin{bmatrix}
    \mathbf{s}^\mathbf{T}_{11}(n) & \cdots & \mathbf{s}^\mathbf{T}_{1K}(n)\\
    \mathbf{s}^\mathbf{T}_{21}(n) & \cdots & \mathbf{s}^\mathbf{T}_{2K}(n)\\
    \vdots & \ddots & \vdots \\
    \mathbf{s}^\mathbf{T}_{M1}(n) & \cdots & \mathbf{s}^\mathbf{T}_{MK}(n)\\
    \end{bmatrix}\in\mathbb{R}^{M\times KL}.
\end{equation}
In the above matrix, $\mathbf{s}_{mk}(n)$ represents the impulse response of the secondary path from the $k$-th secondary source to the $m$-th error sensor:
\begin{equation}
    \mathbf{s}_{mk}(n) =\begin{bmatrix}
    s^{(1)}_{mk}(n) &  s^{(2)}_{mk}(n) & \cdots &  s^{(L)}_{mk}(n) 
    \end{bmatrix}^\mathrm{T}\in\mathbb{R}^{L\times 1}.
\end{equation}

According to above analysis of forward propagation, it is found that the control filter $\mathbf{W}(n)$ will contribute to the anti-noises $\mathbf{y}^\prime(n)$, $\mathbf{y}^\prime(n+1)$, $\cdots$, and $\mathbf{y}^\prime(n+L-1)$. Thus, updating $\mathbf{W}(n)$ requires future gradient knowledge, which violates the causal restriction in real ANC systems.

\subsection{Back propagation}
\begin{figure}
    \centering
    \includegraphics[width=13cm]{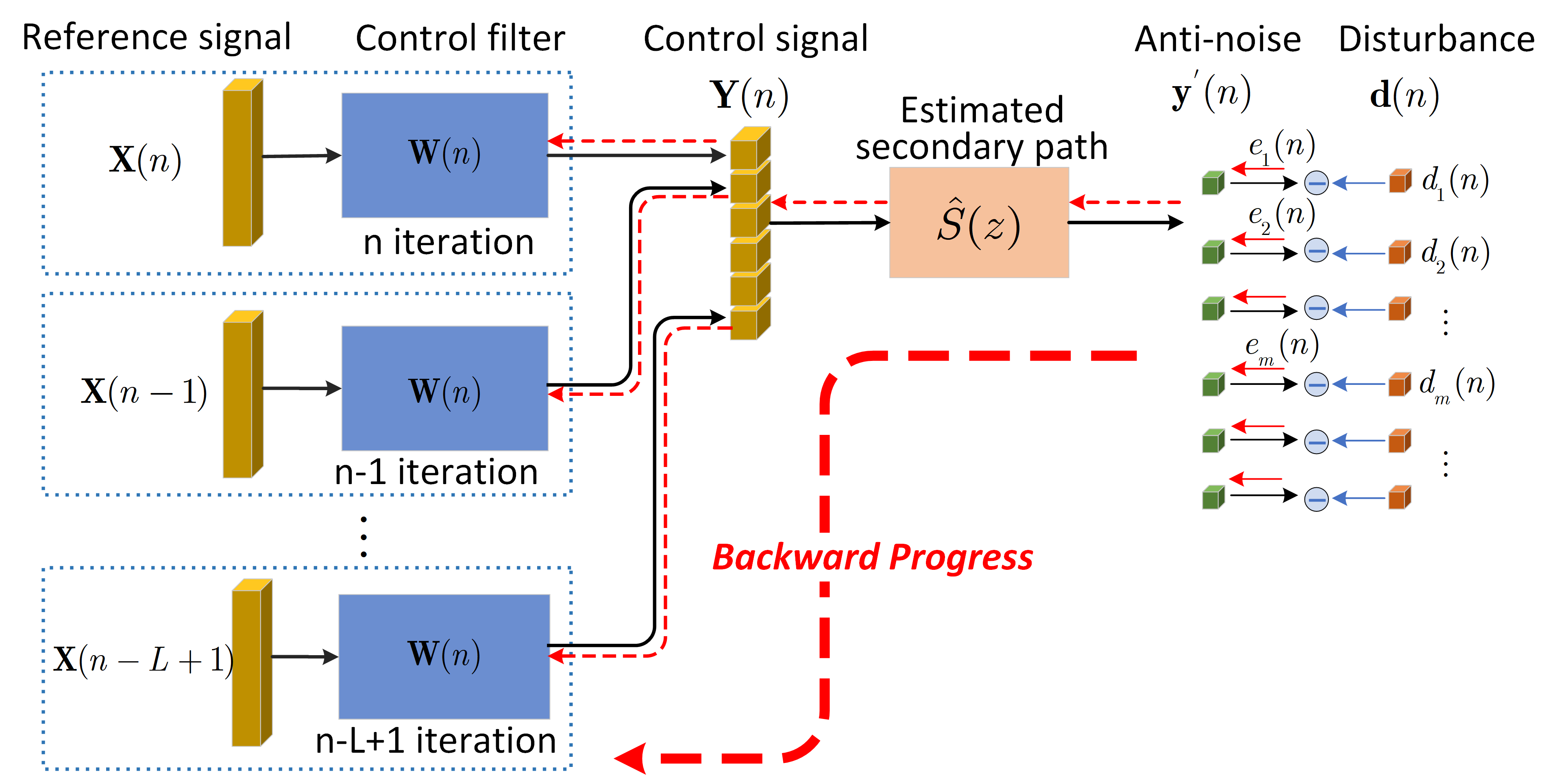}
    \caption{The back propagation of the McFxLMS algorithm.}
    \label{Fig 3}
\end{figure}
To tackle the causal problem during updating, we assumed that the adaptive filtering is a slowly changing process, which means that
\begin{equation}
    \mathbf{W}(n)\approx \mathbf{W}(n-1)\approx \dots \approx \mathbf{W}(n-L+1).
\end{equation}
Therefore, the forward propagation of the McFxLMS algorithm can be modified as shown in Figure \ref{Fig 3}. The $k$-th control signal is obtained from
\begin{equation}
    y_k(n)=\sum^{J}_{j=1}\mathbf{w}^\mathrm{T}_{kj}\mathbf{x}_j(n),~~~j=1,\cdots,J~\text{and}~k=1,\cdots,K
\end{equation}
and its vector format can be rewritten as
\begin{equation}
    \mathbf{y}_k(n) =\begin{bmatrix}
    \sum^{J}_{j=1}\mathbf{w}^\mathrm{T}_{kj}\mathbf{x}_j(n) & \sum^{J}_{j=1}\mathbf{w}^\mathrm{T}_{kj}\mathbf{x}_j(n-1) & \cdots &
    \sum^{J}_{j=1}\mathbf{w}^\mathrm{T}_{kj}\mathbf{x}_j(n-L+1)
    \end{bmatrix}^\mathrm{T}\in\mathbb{R}^{L\times 1}.
\end{equation}
Hence, the $m$-th anti-noise can be calculated as
\begin{equation}
    y^\prime_m(n)=\sum^{K}_{k=1}\hat{\mathbf{s}}^\mathrm{T}_{mk}\mathbf{y}_k(n)~~~m=1,\cdots,M
\end{equation}
where $\hat{\mathbf{s}}_{mk}$ denotes the estimate of the secondary path $\mathbf{s}_{mk}(n)$.

Moreover, the cost function of the McFxLMS algorithm is defined as the sum of squared error signal on each error sensor:
\begin{equation}
    \mathbf{J} =\sum^{M}_m e^2_m =\sum^{M}_m \left[d_m(n)-y^\prime_m(n)\right]^2
\end{equation}
where $d_m(n)$ is the disturbance picked up by the $m$-th error sensor.

\begin{figure}
    \centering
    \includegraphics{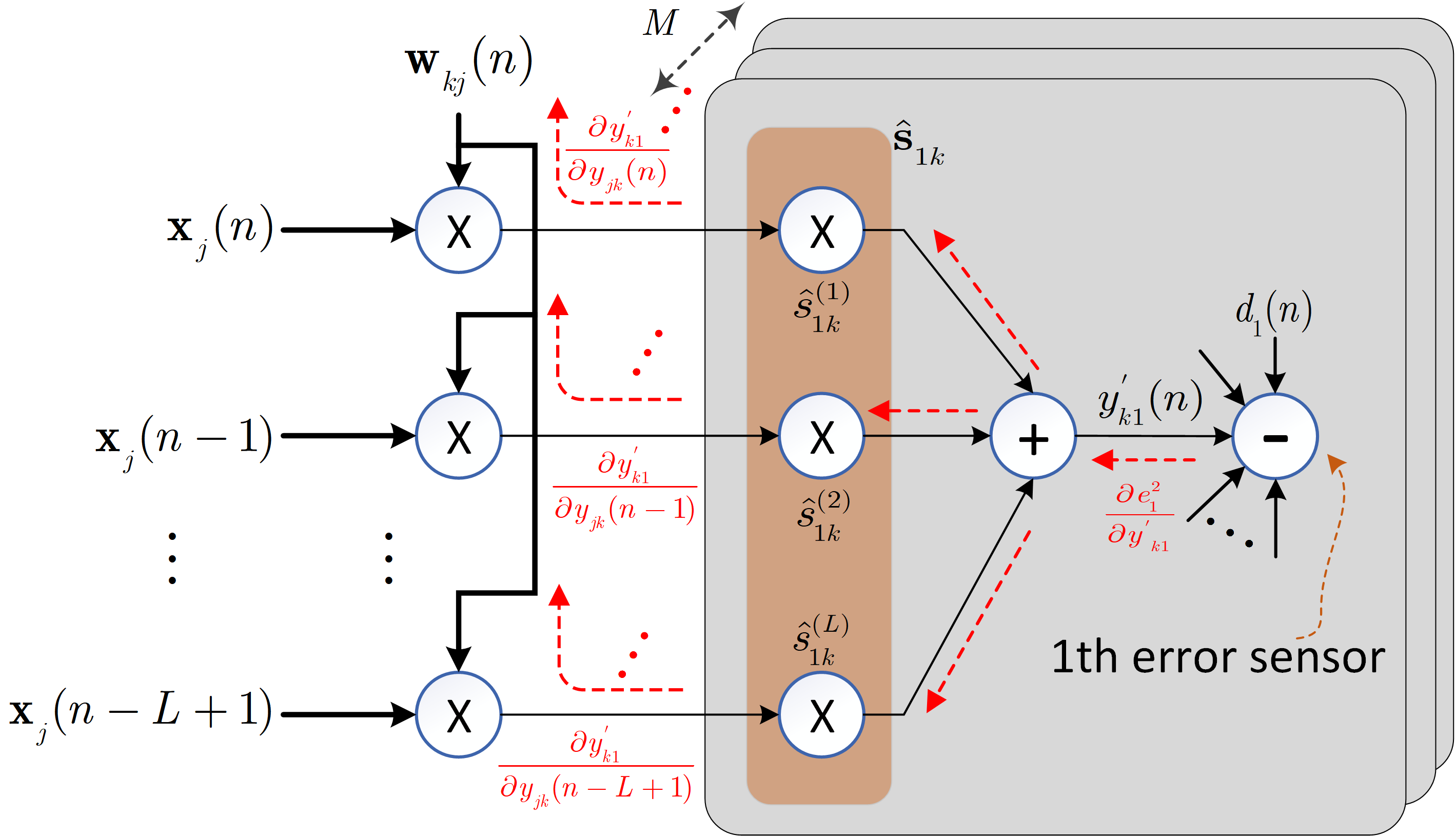}
    \caption{The computing graph of the control filter $\mathbf{w}_{kj}(n)$ in the McFxLMS algorithm under the slow adapting assumption, where $j=1,\cdots, J$ and $k=1, \cdots, M$.}
    \label{Fig 4}
\end{figure}

Through the above discussion, we can get the updating graph of the control filter $\mathbf{w}_{kj}(n)$ as shown in Figure \ref{Fig 4}. According to back propagation mechanism and chain derivation rule, the gradient with respect to $\mathbf{w}_{kj}$ can be calculated by
\begin{equation}\label{eq_18}
    \begin{split}
        \nabla_{\mathbf{w}_{kj}}&=\frac{\partial\mathbf{J}}{\partial \mathbf{w}_{kj}}=\frac{\partial\mathbf{y}_k(n)}{\partial \mathbf{w}_{kj}}\sum^{M}_{m=1}\frac{\partial e^2_m(n)}{\partial y^\prime_m(n)}\frac{\partial y^\prime_m(n)}{\partial\mathbf{y}_k(n)}\\
        &=-\begin{bmatrix}\mathbf{x}_j(n) & \cdots & \mathbf{x}_j(n-L+1)\end{bmatrix}\sum^{M}_{m=1}2e_m(n)\hat{\mathbf{s}}_{mk}\\
        &=-2\sum^M_{m=1}e_m(n)\mathbf{x}^\prime_{jkm}(n),
    \end{split}
\end{equation}
where the filtered reference signal vector is given by 
\begin{equation}
    \mathbf{x}^\prime_{jkm}(n) = \begin{bmatrix}\mathbf{x}_j(n) & \cdots & \mathbf{x}_j(n-L+1)\end{bmatrix} \hat{\mathbf{s}}_{mk}. 
\end{equation}

Finally, by using the negative value of the gradient \eqref{eq_18} to update the control filer, we can get
\begin{equation}
\mathbf{w}_{kj}(n+1)=\mathbf{w}_{kj}(n)-\frac{\mu}{2}\nabla_{\mathbf{w}_{kj}}=\mathbf{w}_{kj}(n)+\mu\sum^M_{m=1}e_m(n)\mathbf{x}^\prime_{jkm}(n)~~~k=1,\cdots,K~\text{and}~j=1,\cdots,J
\end{equation}
where $\mu$ denotes the step size.

After the analysis of forward and back propagation, we found that the adaptive ANC algorithm may be implemented via automatic derivation mechanism. Since the automatic derivation mechanism in PyTorch can update weights given the cost function, the back propagation process can be automatically achieved without resorting to extra-human efforts. Also, existing optimizers in PyTorch tool like SGD, Adam etc. can be directly employed to update the control filters. Owing to the simple implementation method, changing the number of microphones and loudspeakers in multiple-channel ANC systems is more convenient, so that the practicality is enhanced. Additionally, based on the PyTorch tool, it is convenient to add constraints on the control filter coefficients, such as $L1$ and $L2$ normalization.

\begin{figure}[htbp]
\setlength{\abovecaptionskip}{0.cm}
\setlength{\belowcaptionskip}{-0.cm}
\centering
\subfigure{
\includegraphics[width=0.3\linewidth]{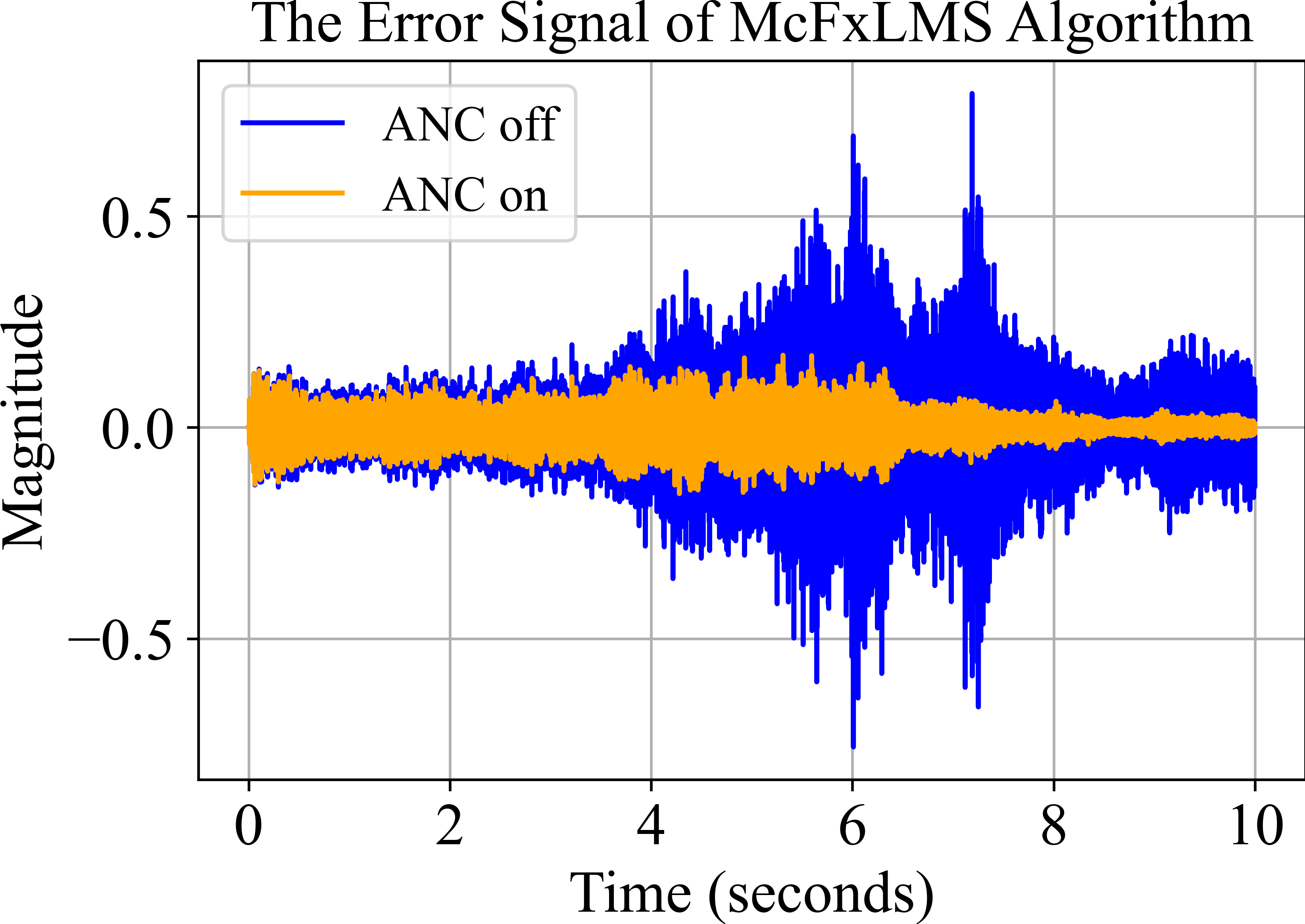}
}
\subfigure{
\includegraphics[width=0.3\linewidth]{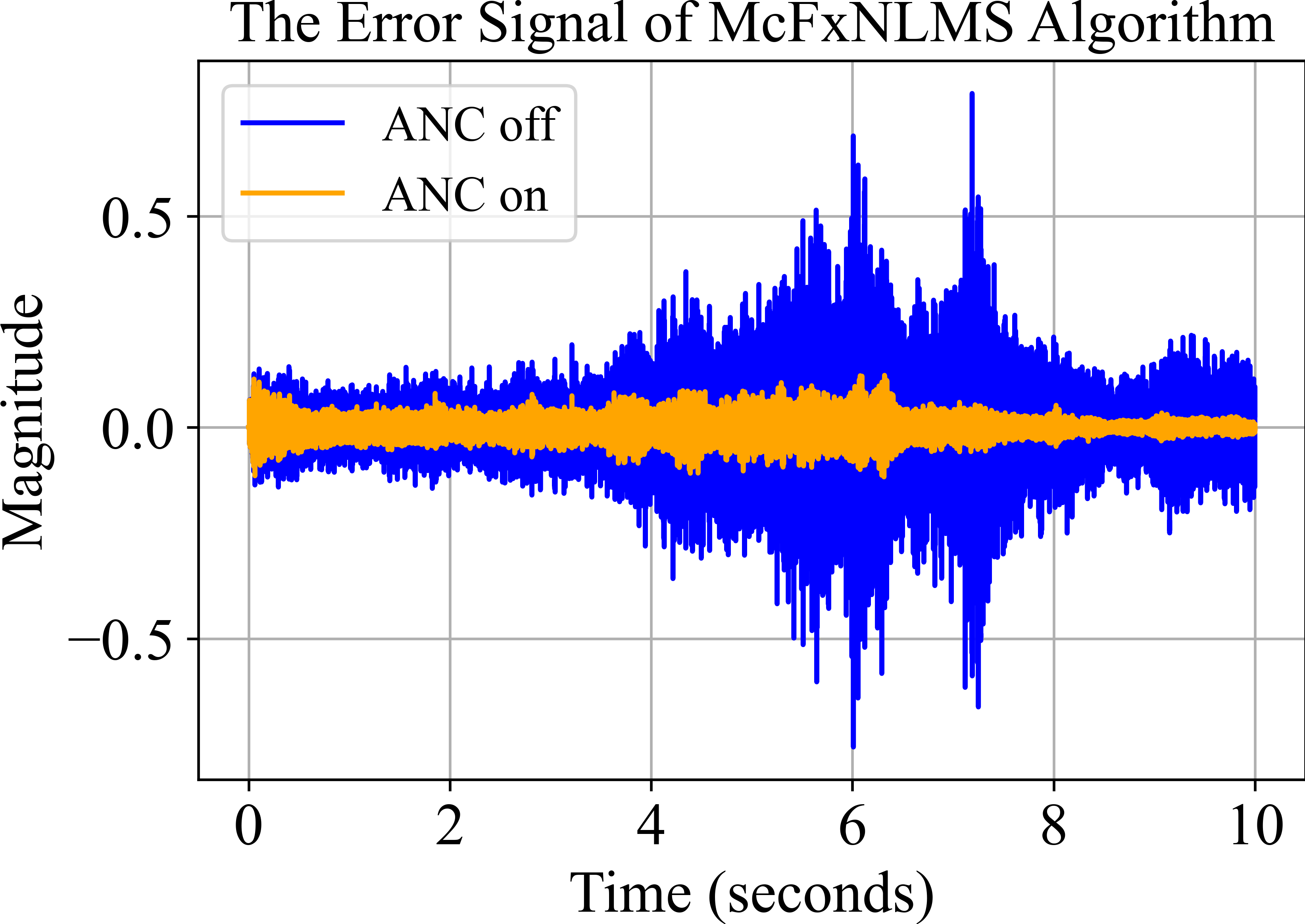}
} \\
\subfigure{
\includegraphics[width=0.3\linewidth]{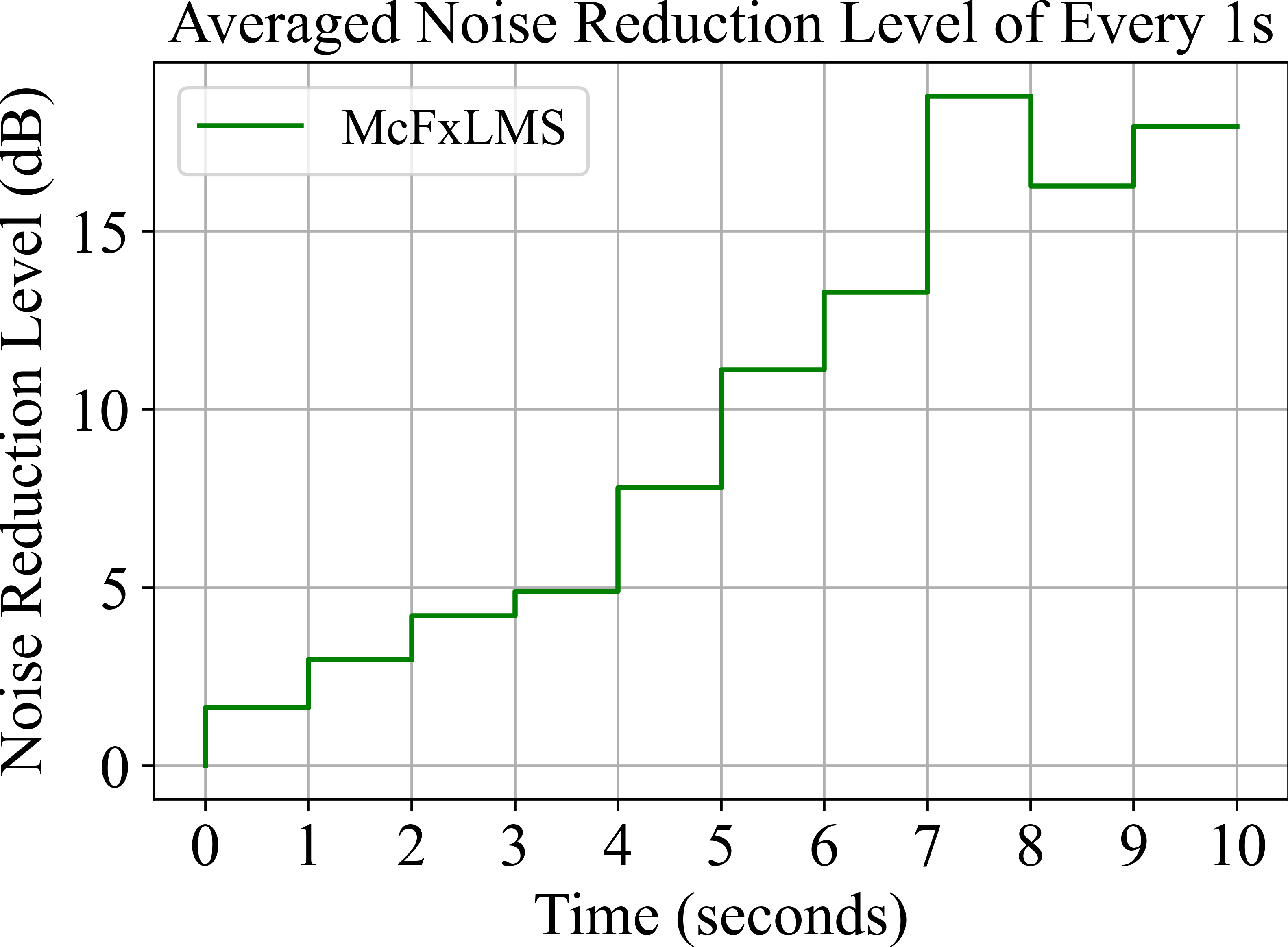}
}
\subfigure{
\includegraphics[width=0.3\linewidth]{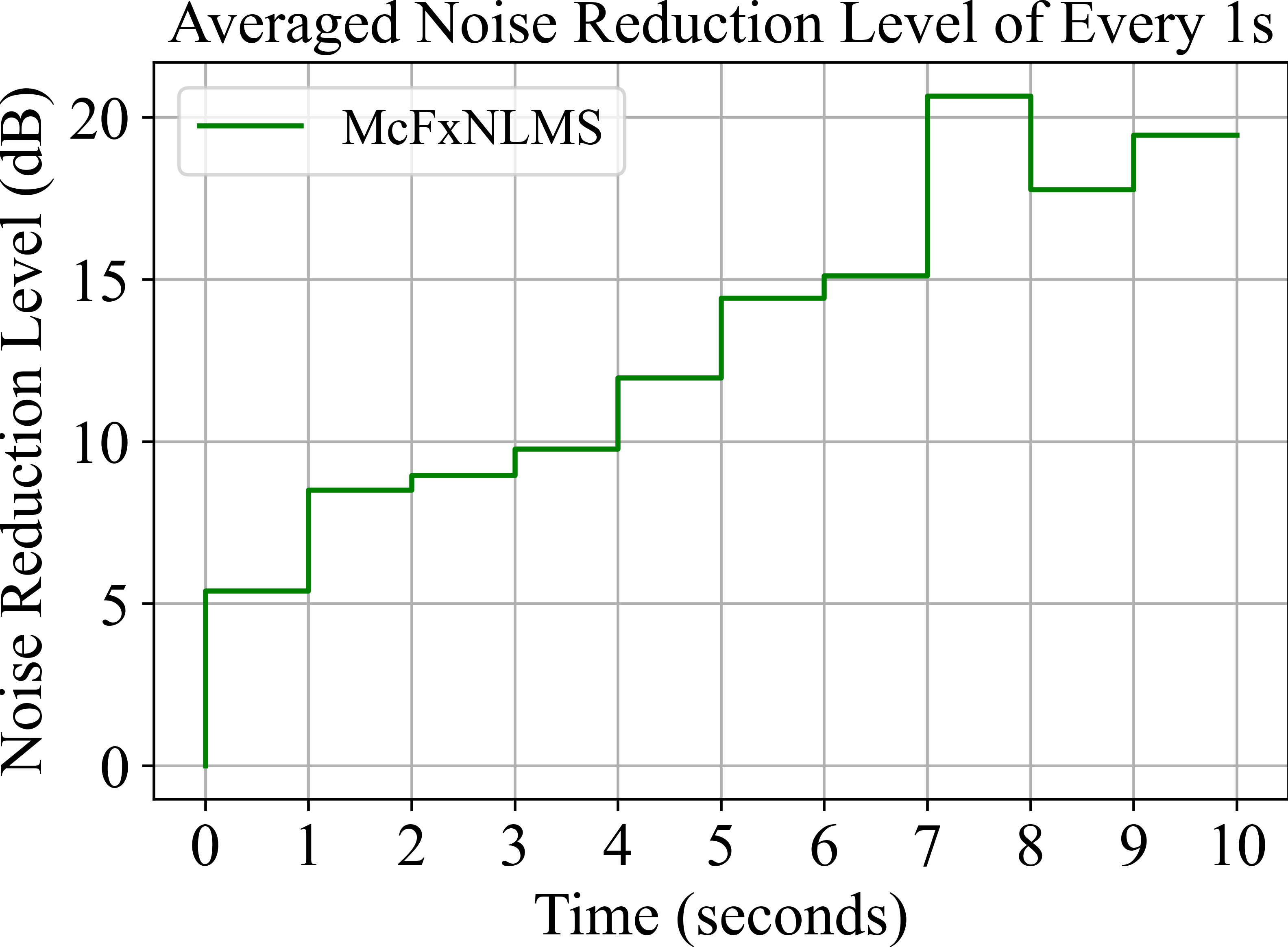}
}
\caption{Error signals and averaged noise reduction level of every 1 second on the aircraft noise.}
\label{Fig 5}\vspace*{-0.5cm}
\end{figure}

\begin{figure}[htbp]
\setlength{\abovecaptionskip}{0.cm}
\setlength{\belowcaptionskip}{-0.cm}
\centering
\subfigure{
\includegraphics[width=0.3\linewidth]{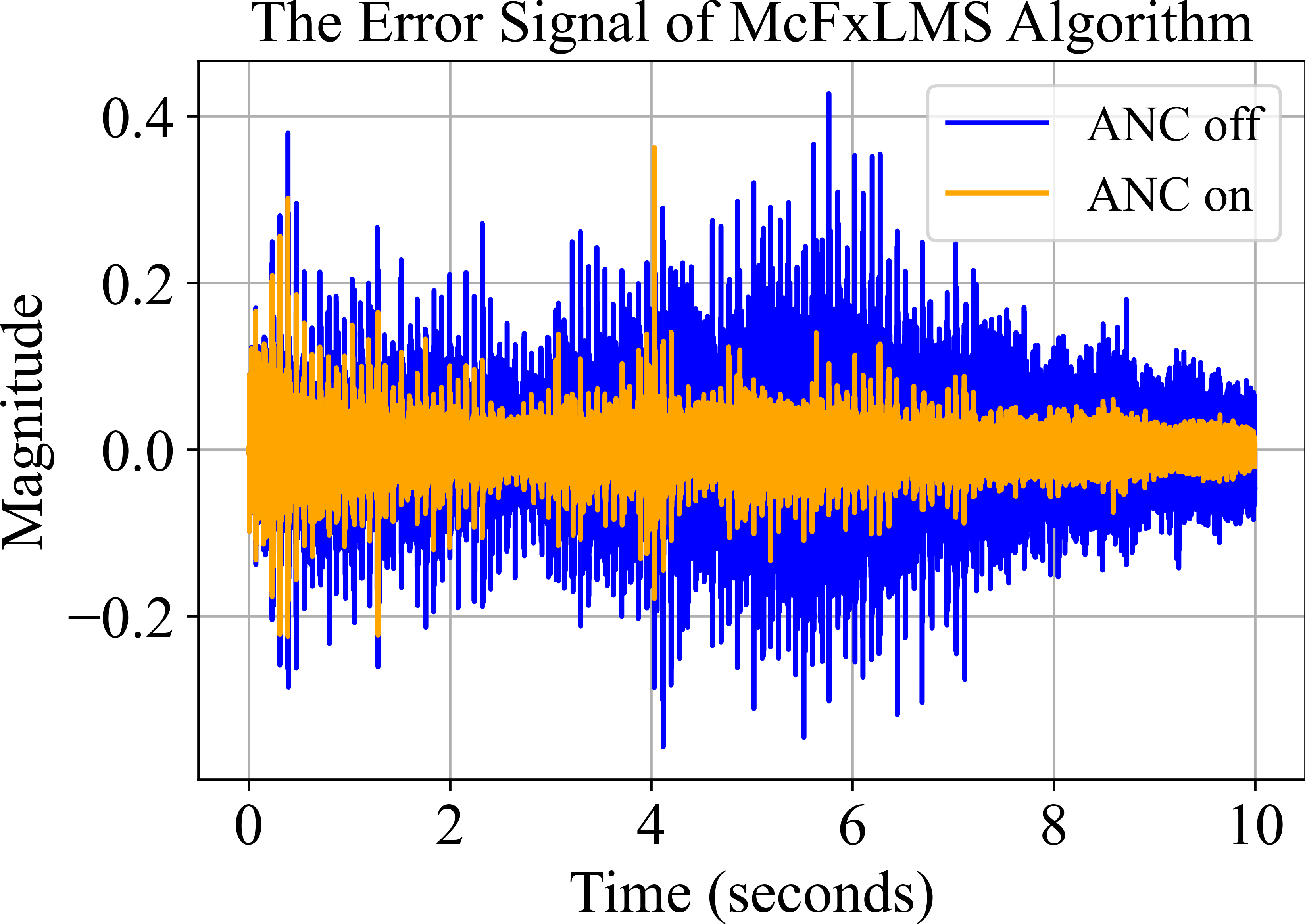}
}
\subfigure{
\includegraphics[width=0.3\linewidth]{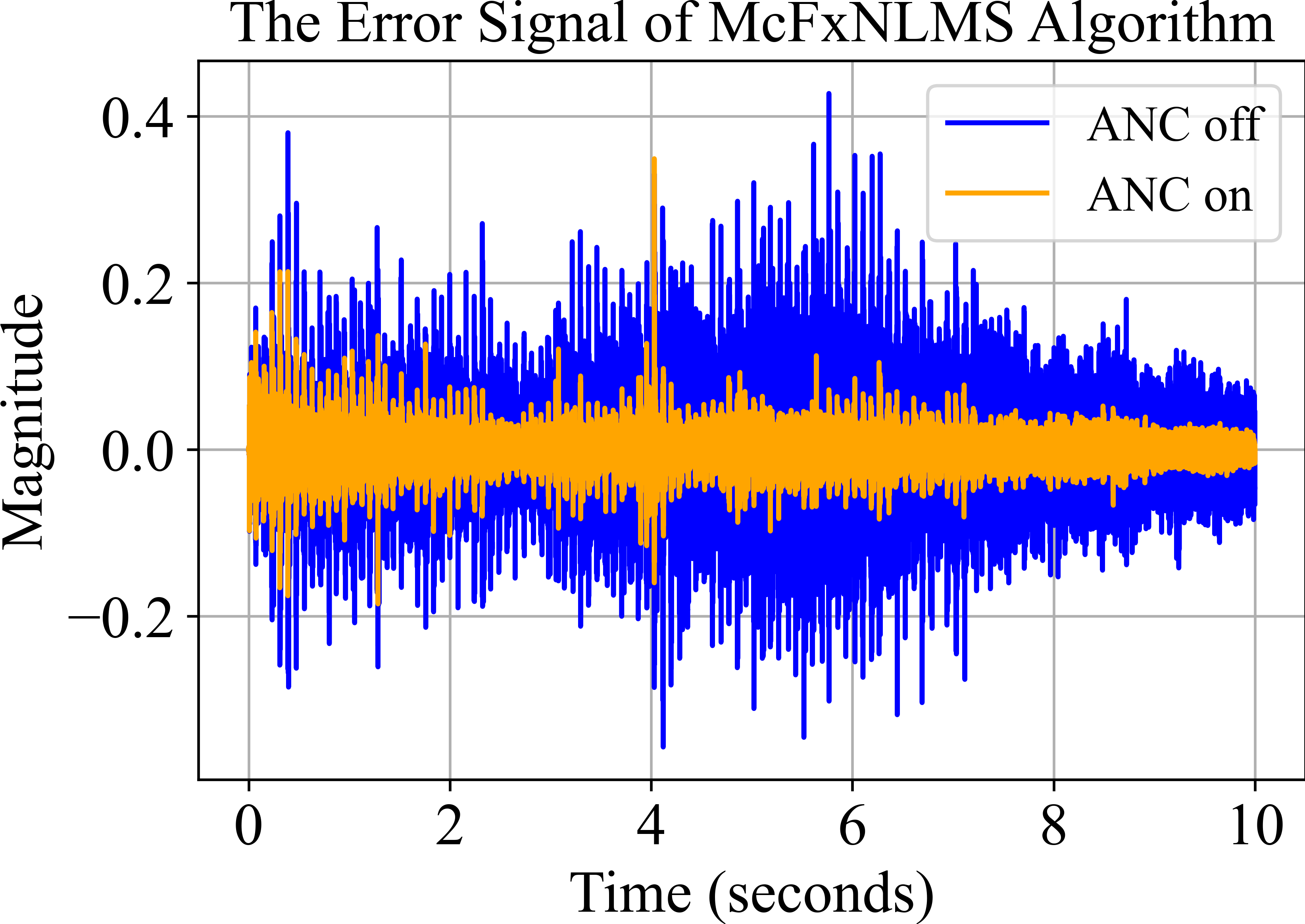}
} \\
\subfigure{
\includegraphics[width=0.3\linewidth]{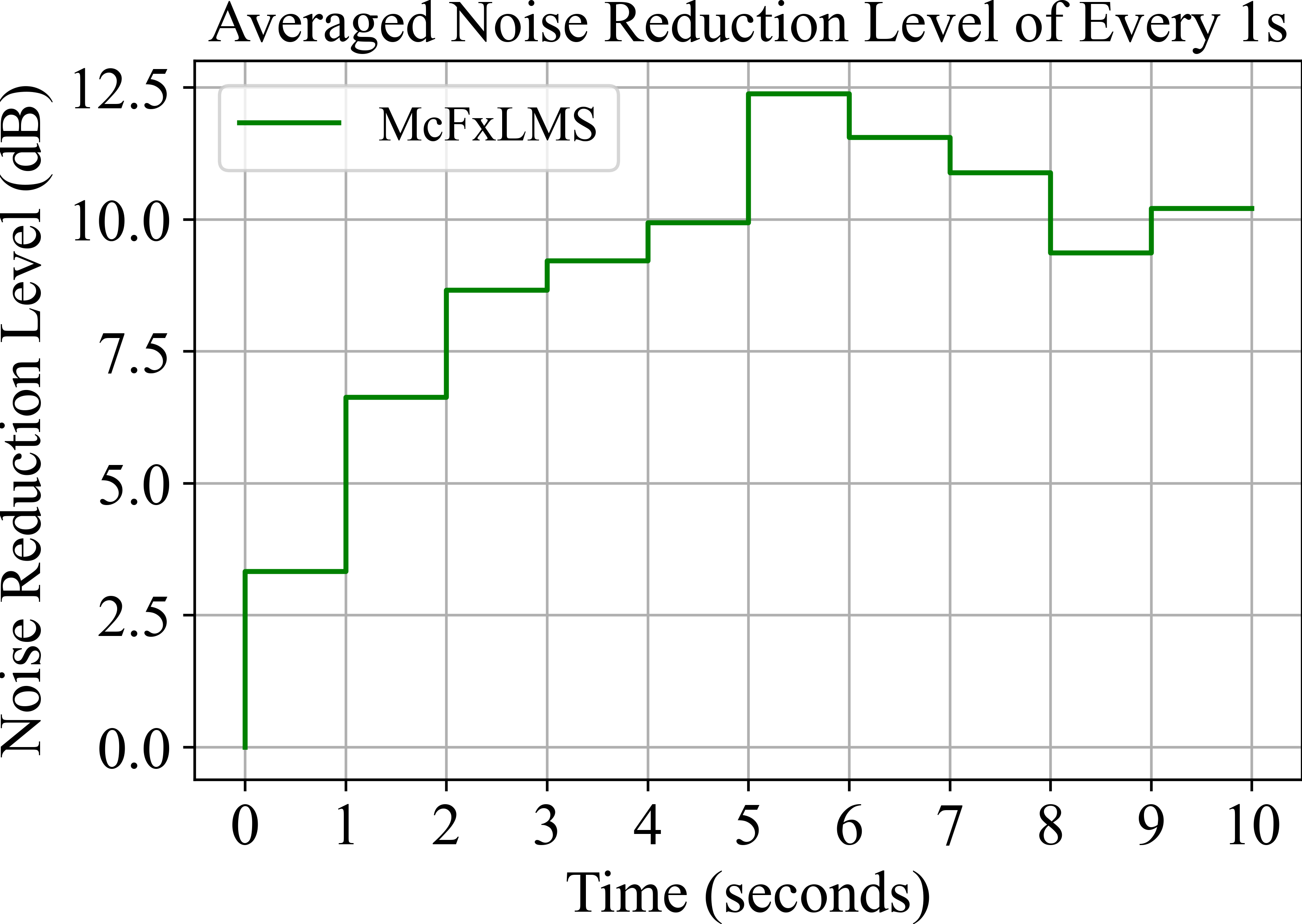}
}
\subfigure{
\includegraphics[width=0.3\linewidth]{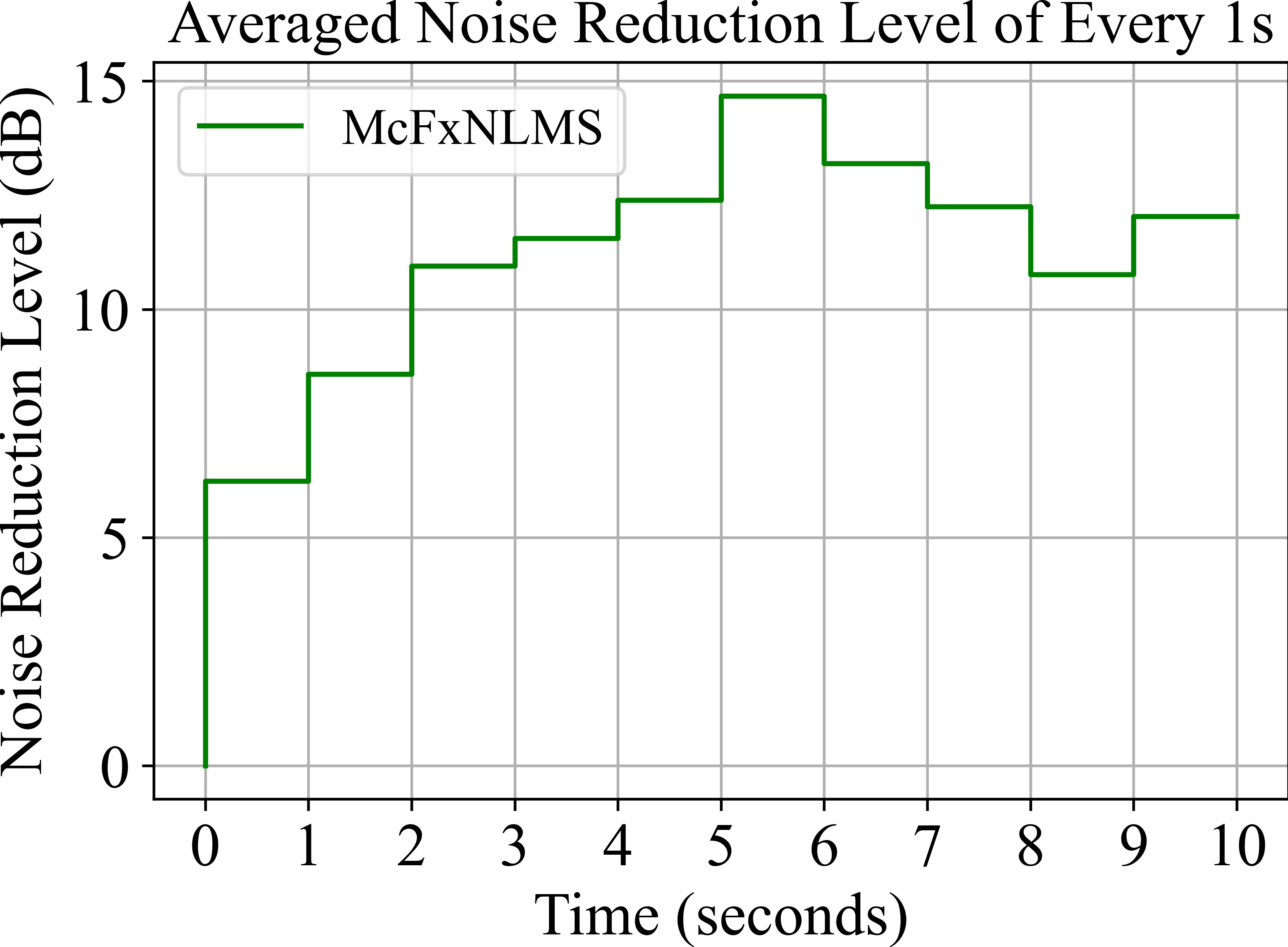}
}
\caption{Error signals and averaged noise reduction level of every 1 second on the helicopter noise.}
\label{Fig 6}\vspace*{-0.5cm}
\end{figure}

\section{\uppercase{Simulation Results}}
We implemented the McFxLMS and McFxNLMS algorithm based on the automatic derivation mechanism in PyTorch. In the ANC system, the number of reference sensors, secondary sources and error sensors are 4, 4, and 4, respectively. The control filter length is $512$ taps. The step size of the McFxLMS algorithm and McFxNLMS algorithm are set to $0.00001$ and $0.001$ separately. The primary path and secondary path are chosen as a band-pass filter and a low-pass filter.

The implemented McFxLMS algorithm and McFxNLMS algorithm are evaluated to suppress several real-recorded noises including: an aircraft noise with a frequency range of $50$Hz-$12,000$Hz, a helicopter noise with a frequency range of $50$Hz-$9,800$Hz, and a traffic noise with a frequency range of $40$Hz-$1,400$Hz. In this section, the stochastic gradient descent (SGD) algorithm \cite{18} was used for optimization.

\begin{figure}[htbp]
\setlength{\abovecaptionskip}{0.cm}
\setlength{\belowcaptionskip}{-0.cm}
\centering
\subfigure{
\includegraphics[width=0.3\linewidth]{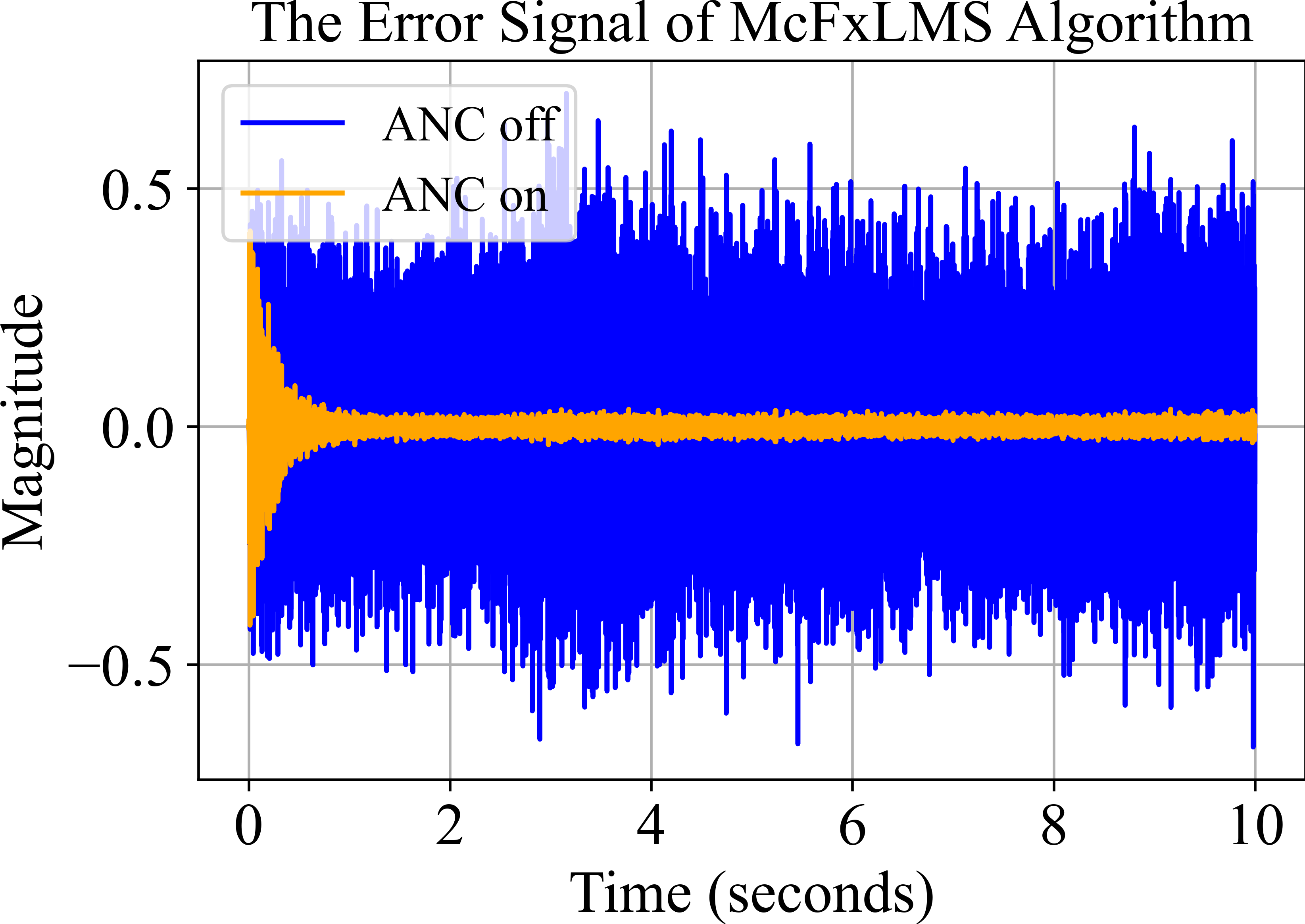}
}
\subfigure{
\includegraphics[width=0.3\linewidth]{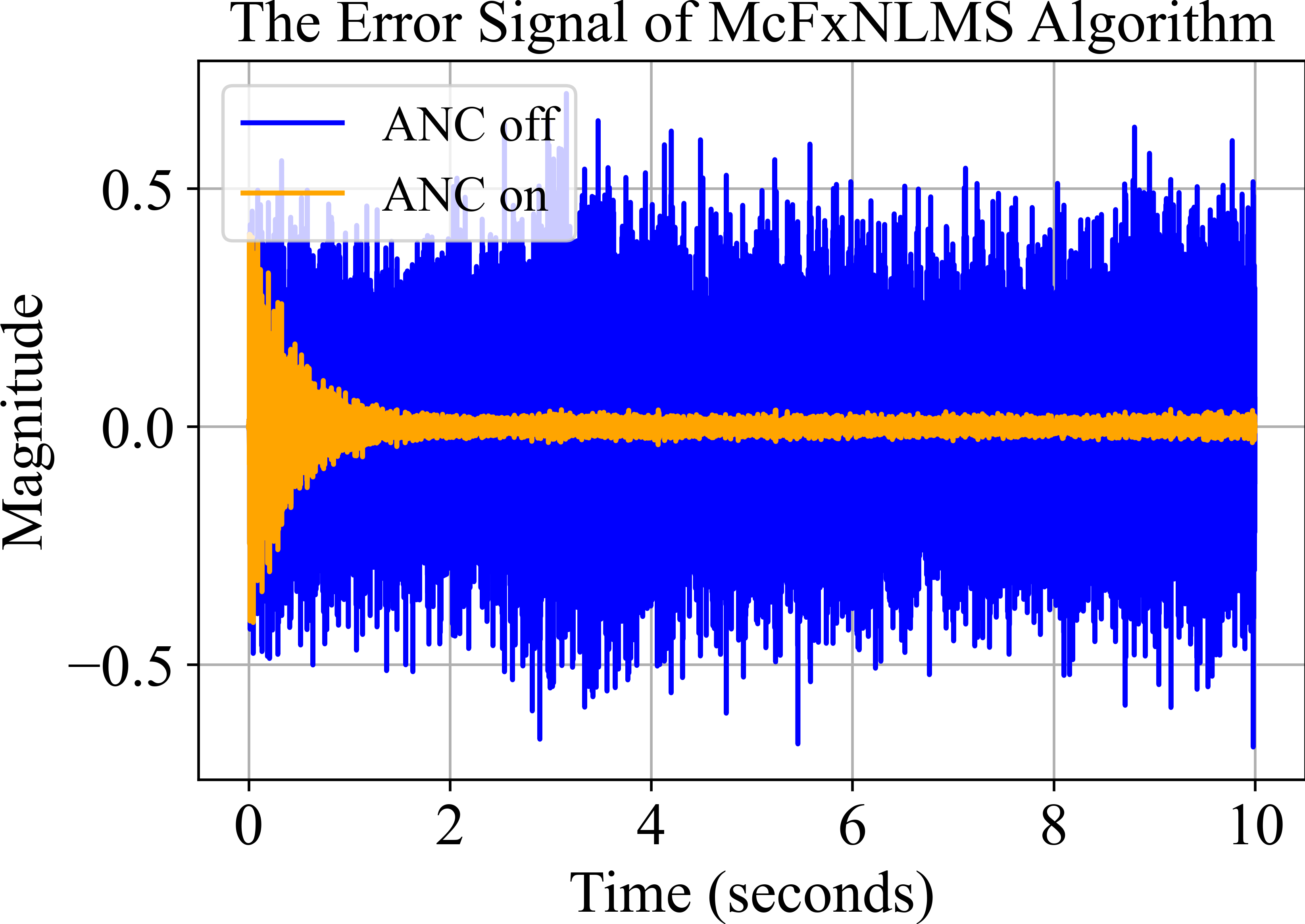}
} \\
\subfigure{
\includegraphics[width=0.3\linewidth]{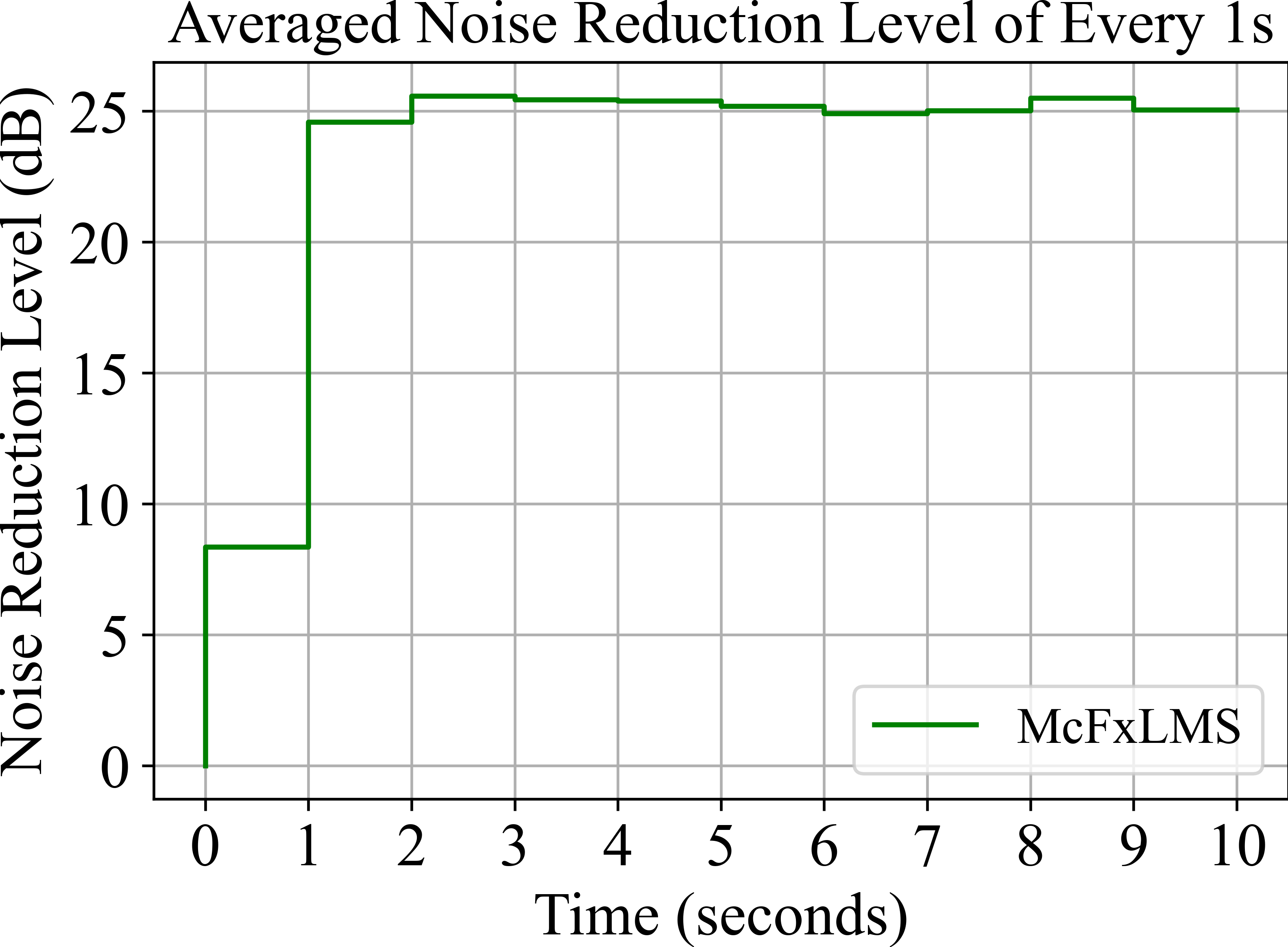}
}
\subfigure{
\includegraphics[width=0.3\linewidth]{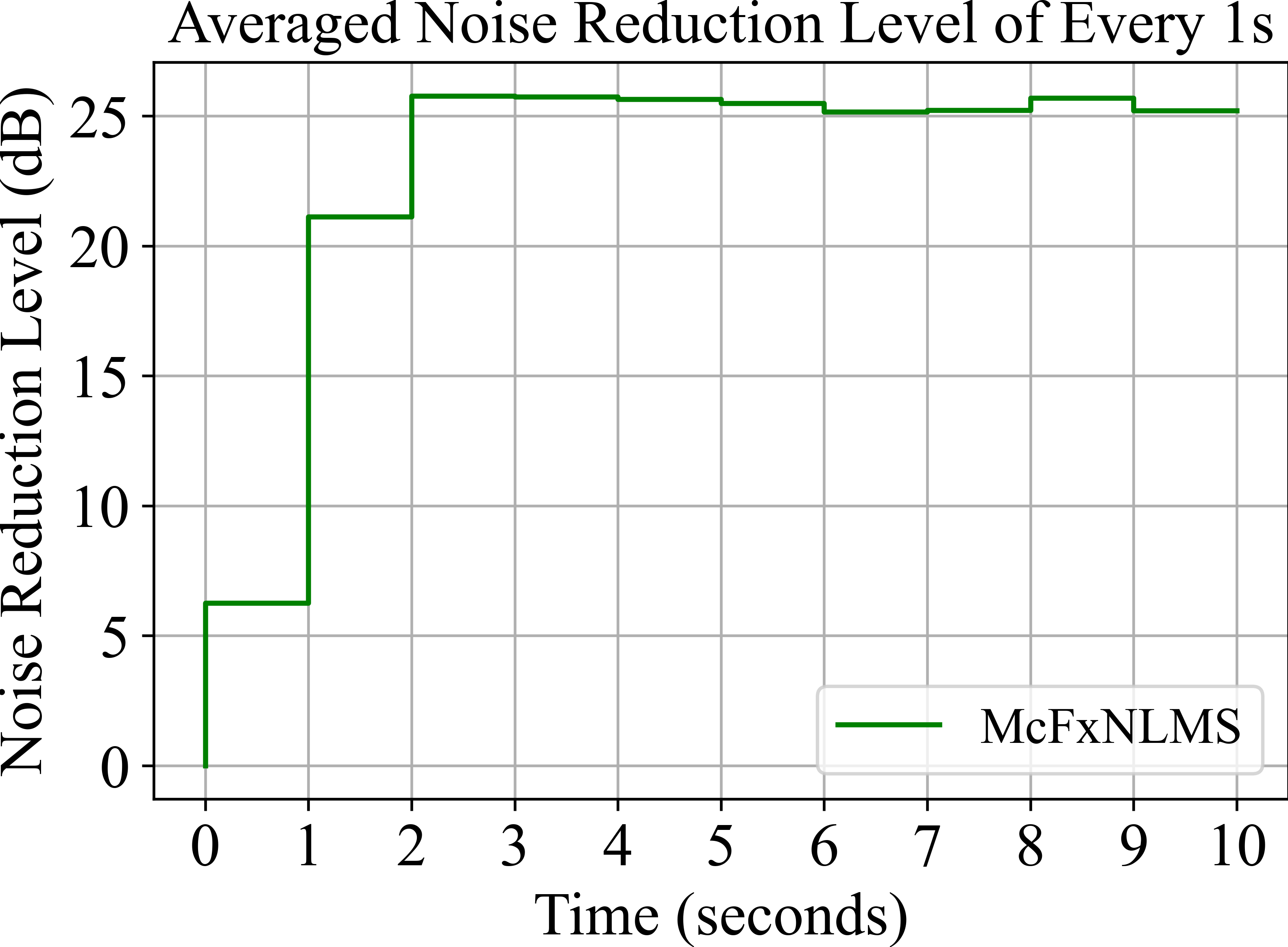}
}
\caption{Error signals and averaged noise reduction level of every 1 second on the traffic noise.}
\label{Fig 7}\vspace*{-0.5cm}
\end{figure}

The noise reduction results using the implemented McFxLMS algorithm and McFxNLMS algorithm on the aircraft noise, helicopter noise, and traffic noise are shown in Figure \ref{Fig 5}, Figure \ref{Fig 6}, and Figure \ref{Fig 7}. According to the simulation results, the implemented McFxLMS and McFxNLMS algorithm can attenuate the real-world noises, which demonstrates the effectiveness of the implementation method. Also, compared with the McFxLMS algorithm, the McFxNLMS algorithm obtains slightly higher steady-state noise reduction levels on the aircraft noise and helicopter noise. However, the McFxNLMS algorithm responds more slowly to the traffic noise than the McFxLMS algorithm.

\section{\uppercase{Conclusions}}
In this paper, we implement the McFxLMS and McFxNLMS algorithm based on the back propagation mechanism. Since the automatic derivation in PyTorch can update weights given the cost function, the back propagation of McFxLMS and McFxNLMS can be automatically achieved without resorting to extra-human efforts. Therefore, the simple implementation method of the McFxLMS and McFxNLMS algorithm can improve the practicality. Also, we provide the open source code of the implementation method. Simulation results show that the implemented McFxLMS and McFxNLMS algorithm are effective to attenuate different real noises. Furthermore, this method is expected to implement other adaptive algorithms in multi-channel ANC systems.

\section*{\uppercase{Acknowledgements}}
This research is supported by the Singapore Ministry of National Development and the National Research Foundation, Prime Minister’s Office under the Cities of Tomorrow (CoT) Research Programme (CoT Award No. COT-V4-2019-1). Any opinions, findings, and conclusions or recommendations expressed in this material are those of the author(s) and do not reflect the views of the Singapore Ministry of National Development and National Research Foundation, Prime Minister's Office, Singapore.

\bibliographystyle{vancouver} 
\renewcommand{\refname}{\normalfont\selectfont\normalsize}
\noindent \section*{\uppercase{References}}
\vspace{-18pt}
\bibliography{reference}

\begin{thebibliography}{10}

\bibitem{1}
Hansen CN.
\newblock Understanding active noise cancellation.
\newblock CRC Press; 2002.

\bibitem{2}
Elliott SJ, Nelson PA.
\newblock Active noise control.
\newblock IEEE signal processing magazine. 1993;10(4):12-35.

\bibitem{3}
Kuo SM, Morgan DR.
\newblock Active noise control: a tutorial review.
\newblock Proceedings of the IEEE. 1999;87(6):943-73.

\bibitem{13}
Shi C, Xie R, Jiang N, Li H, Kajikawa Y.
\newblock Selective virtual sensing technique for multi-channel feedforward
  active noise control systems.
\newblock In: ICASSP 2019-2019 IEEE International Conference on Acoustics,
  Speech and Signal Processing (ICASSP). IEEE; 2019. p. 8489-93.

\bibitem{4}
Zhang H, Wang D.
\newblock A Deep Learning Method to Multi-Channel Active Noise Control.
\newblock In: Interspeech; 2021. p. 681-5.

\bibitem{5}
Jung TH, Kim JH, Kim KJ, Nam SW.
\newblock Active noise reduction system using multi-channel ANC.
\newblock In: 2011 11th International Conference on Control, Automation and
  Systems. IEEE; 2011. p. 36-9.

\bibitem{6}
Oh SH, Kim Hs, Park Y.
\newblock Active control of road booming noise in automotive interiors.
\newblock The Journal of the Acoustical Society of America. 2002;111(1):180-8.

\bibitem{7}
Kuo SM, Finn BM.
\newblock A general multi-channel filtered LMS algorithm for 3-D active noise
  control systems.
\newblock Second Int Con on Recent Developments in Air and Structure Borne
  Sound and Vibration, Auburn AL. 1992:345-52.

\bibitem{8}
Elliot S, Nelson P, Stothers I, Boucher C.
\newblock In-flight experiments on the active control of propeller-induced
  cabin noise.
\newblock Journal of Sound and Vibration. 1990;140(2):219-38.

\bibitem{14}
Shi D, Gan WS, Lam B, Wen S.
\newblock Feedforward selective fixed-filter active noise control: Algorithm
  and implementation.
\newblock IEEE/ACM Transactions on Audio, Speech, and Language Processing.
  2020;28:1479-92.

\bibitem{9}
Garas J.
\newblock Adaptive 3D sound systems. vol. 566.
\newblock Springer Science \& Business Media; 2012.

\bibitem{10}
Elliott S, Stothers I, Nelson P.
\newblock A multiple error LMS algorithm and its application to the active
  control of sound and vibration.
\newblock IEEE Transactions on Acoustics, Speech, and Signal Processing.
  1987;35(10):1423-34.

\bibitem{17}
Kinoshita S, Kajikawa Y.
\newblock Multi-channel feedforward ANC system combined with noise source
  separation.
\newblock In: 2015 Asia-Pacific Signal and Information Processing Association
  Annual Summit and Conference (APSIPA). IEEE; 2015. p. 379-83.

\bibitem{15}
Shi D, Lam B, Ooi K, Shen X, Gan WS.
\newblock Selective fixed-filter active noise control based on convolutional
  neural network.
\newblock Signal Processing. 2022;190:108317.

\bibitem{16}
Luo Z, Shi D, Gan WS.
\newblock A Hybrid SFANC-FxNLMS Algorithm for Active Noise Control Based on
  Deep Learning.
\newblock IEEE Signal Processing Letters. 2022;29:1102-6.

\bibitem{12}
Shi D, Gan WS, Lam B, Wen S, Shen X.
\newblock Active noise control based on the momentum multichannel normalized
  filtered-x least mean square algorithm.
\newblock In: INTER-NOISE and NOISE-CON Congress and Conference Proceedings.
  vol. 261. Institute of Noise Control Engineering; 2020. p. 709-19.

\bibitem{11}
Paszke A, Gross S, Massa F, Lerer A, Bradbury J, Chanan G, et~al.
\newblock Pytorch: An imperative style, high-performance deep learning library.
\newblock Advances in neural information processing systems. 2019;32.

\bibitem{18}
Bottou L.
\newblock Stochastic gradient descent tricks.
\newblock In: Neural networks: Tricks of the trade. Springer; 2012. p. 421-36.

\end{thebibliography}

\end{document}